\documentclass{aa}
\usepackage{graphicx}
\usepackage{txfonts}
\usepackage{natbib}
\bibpunct{(}{)}{;}{a}{}{,} 
\begin{document}
\title{Herschel imaging of the dust in the Helix Nebula (NGC
  7293)\thanks{Herschel is an ESA space observatory with science
    instruments provided by European-led Principal Investigator
    consortia and with important participation from NASA.}}

\author{G.~C.~Van~de~Steene
        \inst{1}\fnmsep\thanks{gsteene@oma.be}
         \and
          P.~A.~M.~van~Hoof \inst{1}
          \and
          K.~M.~Exter\inst{2}
          \and         
         M.~J.~Barlow\inst{3}
  \and
  J.~Cernicharo\inst{10}
  \and
  M.~Etxaluze\inst{10}  
  \and
  W.~K.~Gear\inst{6}
  \and
  J.~R.~Goicoechea\inst{10}
  \and
  H.~L.~Gomez\inst{6}
  \and
  M.~A.~T.~Groenewegen\inst{1}
  \and
  P.~C.~Hargrave\inst{6}
  \and
  R.~J.~Ivison\inst{4}
  \and
  S.~J.~Leeks\inst{7}
  \and
  T.~L.~Lim\inst{7}
  \and
  M.~Matsuura\inst{3}
  \and
  G.~Olofsson\inst{8}
  \and
  E.~T.~Polehampton\inst{7,9}
  \and
  B.~M.~Swinyard\inst{7}
  \and
  T.~Ueta\inst{5}
  \and 
  H.~Van~Winckel\inst{2}
  \and
  C.~Waelkens\inst{2}
  \and
  R.~Wesson\inst{3,11}
}
 
\institute{Royal Observatory of Belgium, Ringlaan 3, B-1180 Brussels, Belgium
  \and
Instituut voor Sterrenkunde, Katholieke Universiteit Leuven, Celestijnenlaan 200 D, B-3001
 Leuven, Belgium
\and 
  Dept of Physics \& Astronomy, University College London, Gower St, London WC1E 6BT, UK
  \and
  UK Astronomy Technology Centre, Royal Observatory Edinburgh, Blackford Hill, Edinburgh EH9 3HJ, UK
  \and
  Dept. of Physics and Astronomy, University of Denver, Mail Stop 6900, Denver, CO 80208, USA
  \and
  School of Physics and Astronomy, Cardiff University, 5 The Parade, Cardiff, Wales CF24 3YB, UK
  \and
  Space Science and Technology Department, Rutherford Appleton Laboratory, Oxfordshire, OX11 0QX, UK
  \and
  Dept.\ of Astronomy, Stockholm University, AlbaNova University Center, Roslagstullsbacken 21, 10691 Stockholm, Sweden
  \and
  Department of Physics, University of Lethbridge, Lethbridge, Alberta, T1J 1B1, Canada
  \and
Instituto de Ciencia de Materiales (ICMM-CSIC), Calle Sor Juana Ines de la Cruz, 3, Cantoblanco, 28049 Madrid, Spain 
\and European Southern Observatory, Alonso de Cordova 3107, Casilla 19001, Santiago, Chile }
   \date{Received 12 May 2014; Accepted 08 December 2014}

  \abstract
  {} 
  {In our series of papers presenting the
  {\em Herschel} imaging of evolved planetary nebulae, we present images of the
  dust distribution in the Helix nebula (NGC~7293).}
  {Images at 70, 160, 250, 350, and 500~$\mu$m were obtained with the PACS and SPIRE instruments on board the Herschel satellite.}
  { The broadband maps show the dust
  distribution over the main Helix nebula to be clumpy and
  predominantly present in the barrel wall. We determined the spectral
  energy distribution of the main nebula in a consistent way using
  Herschel, IRAS, and Planck flux values. The emissivity index of
  $\beta = 0.99 \pm 0.09$, in combination with the carbon rich
  molecular chemistry of the nebula, indicates that the dust consists
  mainly of amorphous carbon. The dust excess emission from the
  central star disk is detected at 70~$\mu$m and the flux measurement
  agrees with previous measurement. We present the temperature and
  dust column density maps. The total dust mass across the Helix
  nebula (without its halo) is determined to be 3.5~$10^{-3}$
  M$_\odot$ at a distance of 216~pc. The temperature map shows dust
  temperatures between 22~K and 42~K, which is similar to the kinetic
  temperature of the molecular gas, confirming that the dust and gas
  co-exist in high density clumps. Archived images are used to compare
  the location of the dust emission in the far infrared ({\em
    Herschel}) with the ionized (GALEX and H$\beta$) and molecular
  (H$_2$) component. The different emission components are consistent
  with the Helix consisting of a thick walled barrel-like structure
  inclined to the line of sight. The radiation field decreases rapidly
  through the barrel wall.  
}
  {}


\maketitle

\section{Introduction}
\label{sectIntro}

We present {\em Herschel} observations of the Planetary Nebula (PN)
NGC~7293, also known as the Helix nebula, which is the nearest large
PN at a distance of 216$^{+14}_{-12}$~pc \citep{Benedict09}, allowing
us to study its spatial structure in detail. The white dwarf central
star WD~2226-210 with a surface temperature of 103~600~$\pm$\,5500~K
\citep{Napiwotzki99} ionizes the AGB nebula. \citet{Su07} also showed
the presence of a 35-150~AU diameter debris disk around this central
star.

The Helix has been extensively studied over a wide range of
wavelengths with ground- and space-based telescopes. At optical
wavelengths neutral and ionized atomic gas has been well-characterized
by the Hubble Space Telescope \citep{ODell04, Meixner05}.
Observations of H$_2$ have identified the presence of numerous
globules often observed with cometary-like tails in H$_2$, CO, and at
infrared wavelengths \citep{Meaburn98, Speck02, Meixner05, Hora06,
  Matsuura09}. A map encompassing the region of the optical image in
the CO J=2$-$1 transition has been made by \citet{Huggins86} and
\citet{Young99}. More complex molecules, including HCN, CN, HCO$^+$,
CCH, C$_3$H$_2$, H$_2$CO, and HCO$^+$ have also been detected in dense
gas in the Helix \citep{Bachiller97, Tenenbaum09, Zack13}. Most
recently \citet{Etxaluze14} studied the spatial distribution of the
atomic and molecular gas along the western rim in the submillimeter
range with the SPIRE instrument on board Herschel.

Understanding the complex morphology of the Helix nebula has been the
subject of numerous studies \citep{Young99, ODell04, Meaburn05}. One
major obstacle has been a lack of high resolution velocity data, until
\citet{Zeigler13} observed the J~=~1$-$0 transition of HCO$^+$ at
89~GHz across the Helix nebula with a velocity resolution of 1.68~km/s
and an angular resolution of 35\arcsec\ half power beamwidth. The
image constructed from the HCO$^+$ emission closely resembles that
observed in the optical, as well as in vibrationally excited H$_2$,
suggesting a common morphology. The high resolution spectra obtained
in HCO$^+$ indicate a barrel-like, bipolar outflow tilted about
10$\degr$ east relative to the line of sight. Around this main nebula
is a 40\arcmin\, diameter halo which shows interaction with the ISM in
the NE \citep{Zhang12, Meaburn13}.

The reason for studying NGC~7293 is its similarity to the other highly
evolved PNe which are also part of the MESS program: the
Ring Nebula (NGC~6720) \citep{vanHoof10} and the Little
Dumbbell Nebula (NGC~650) \citep{vanHoof13}. Both nebulae have a highly
evolved central star and exhibit high density, cometary knots.

In this paper we discuss far-infrared images obtained with the PACS
and SPIRE instruments on board the {\em Herschel} telescope in
Sects.~\ref{sectObs} and \ref{sectMorph}. We present the SED in
Sect.~\ref{sectSED}, the temperature map in Sect.~\ref{sectTemp},
and the mass column density map in Sect.~\ref{sectMass}. The
properties of the central star disk are discussed in Sect.~\ref{sectDisk}. 
In Sect.~\ref{sectComp} we compare the dust
component with images of the molecular and ionized component. The
conclusions are presented in Sect.~\ref{sectConcl}.

\section{Observations and data reduction}
\label{sectObs}

As part of the {\em Herschel} guarantee time key project MESS (Mass loss of
Evolved StarS) (P.I. Martin Groenewegen) we have imaged a sample of
PNe with the PACS \citep{Poglitsch10} and SPIRE
\citep{Griffin10} instruments on board the {\em Herschel} satellite
\citep{Pilbratt10}. A detailed description of the program can be found
in \citet{Groenewegen11}. An overview of the {\em Herschel} observations of
PNe in the MESS program was presented in \citet{vanHoof12}. Other PNe
have been observed in the {\em Herschel} Planetary Nebulae Survey
\citep{Ueta14}. In this article we will present the {\em Herschel} PACS and
SPIRE observations of the Helix nebula obtained in the framework of
the MESS program and the deep SPIRE photometer maps obtained in the
framework of the Must-Do 7 proposal led by J.~Cernicharo.

PACS had three wavelength channels: 70~$\mu$m (blue channel),
100~$\mu$m (green), and 160~$\mu$m (red), with two observed
simultaneously, red and blue in our case. SPIRE operated at
250~$\mu$m (PSW band), 350~$\mu$m (PMW), and 500~$\mu$m (PLW)
simutaneously. We obtained both SPIRE and PACS data simultaneously in
parallel mode at 60\arcsec/s scan speed. SPIRE observations were also
redone as MustDo (MD) observations at a scan speed of 30\arcsec/s.
The log of the observations is presented in Table~\ref{tabObs}.

\begin{figure*}
\begin{center}
\includegraphics[width=\columnwidth,height=0.3\textheight,keepaspectratio]{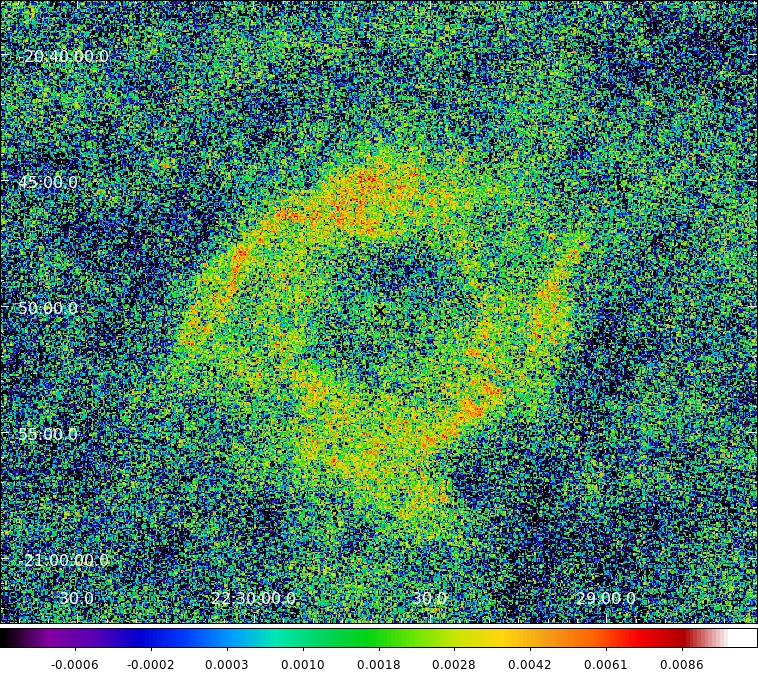} 
\includegraphics[width=\columnwidth,height=0.3\textheight,keepaspectratio]{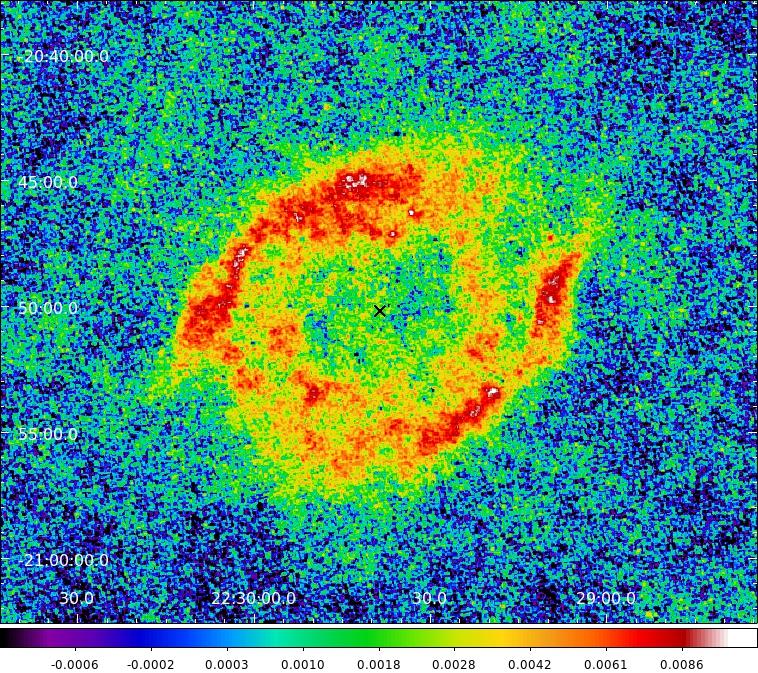}
\includegraphics[width=\columnwidth,height=0.3\textheight,keepaspectratio]{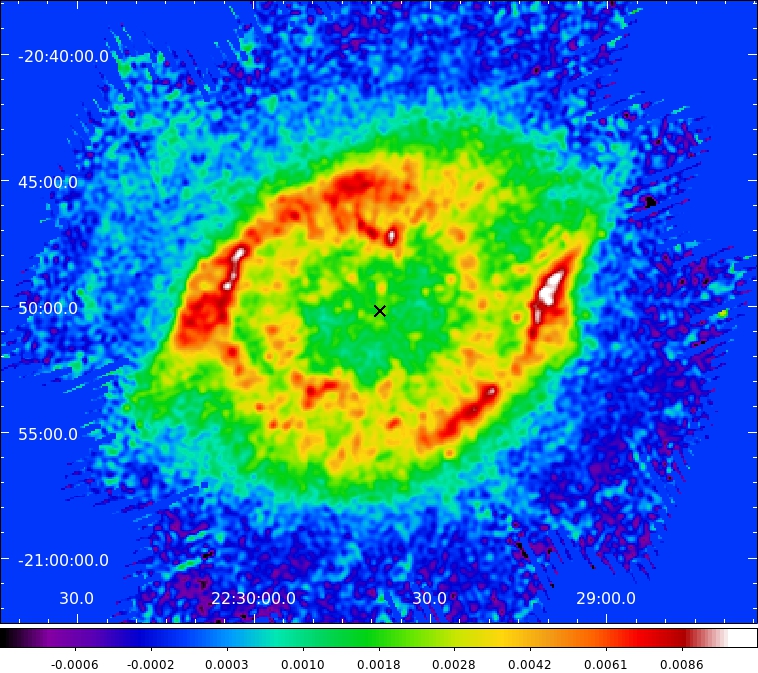}
\includegraphics[width=\columnwidth,height=0.3\textheight,keepaspectratio]{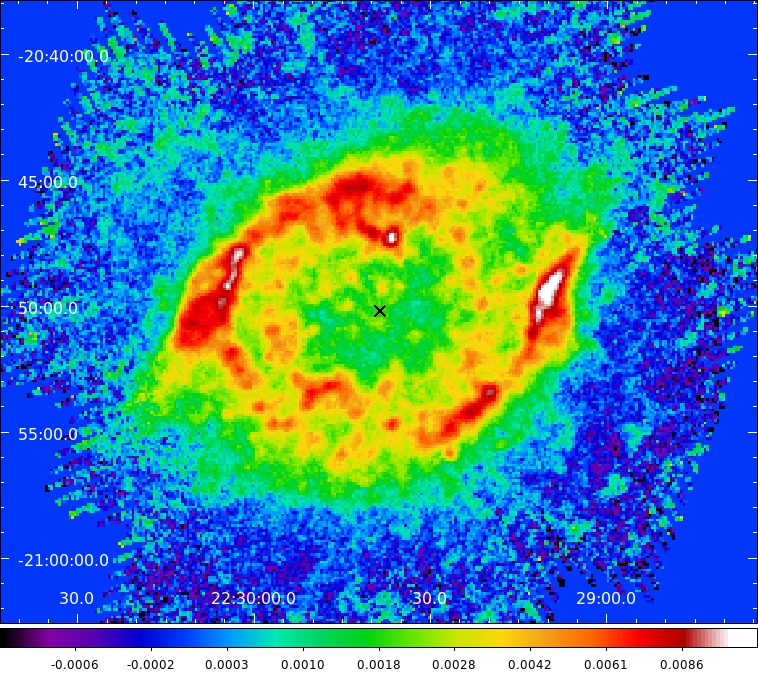} 
\includegraphics[width=\columnwidth,height=0.3\textheight,keepaspectratio]{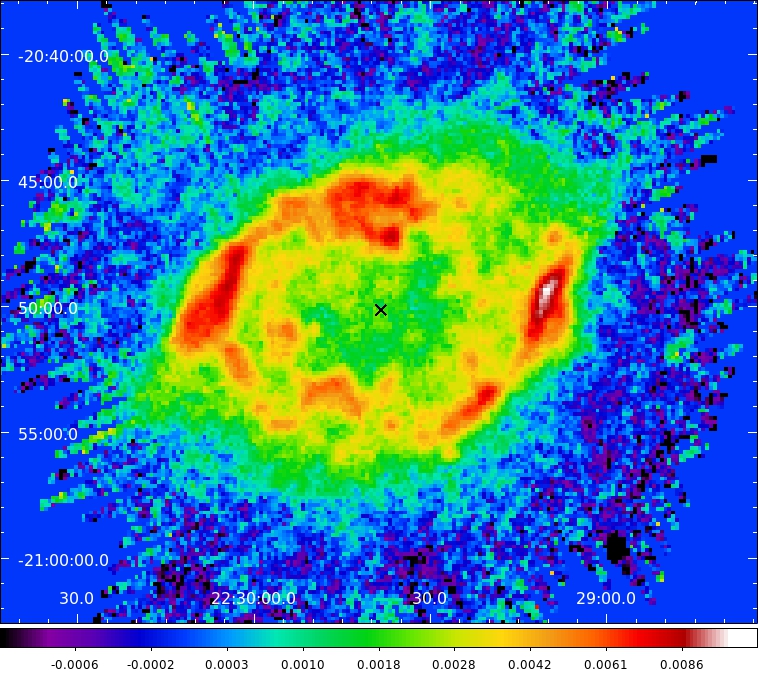}
\includegraphics[width=0.925\columnwidth,keepaspectratio]{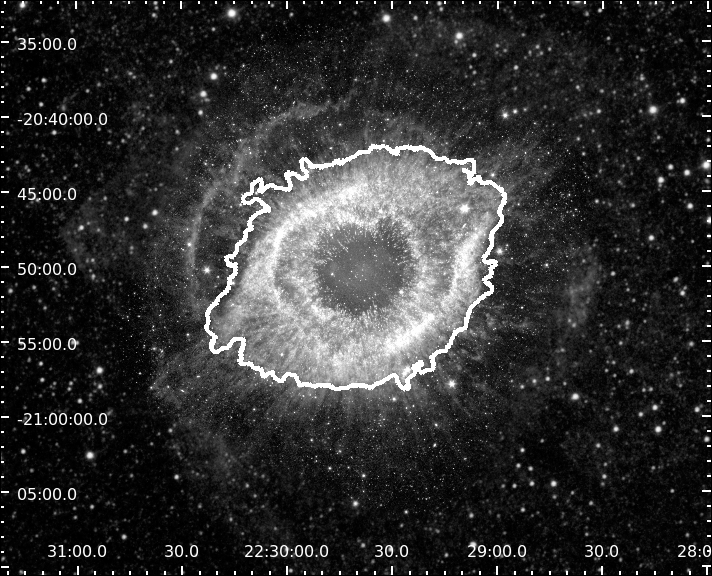}
\caption{ {\em Herschel} images of NGC~7293. Top row left to right show the
  PACS~70 and PACS~160~$\mu$m images, the second row the SPIRE~250 and
  SPIRE~350~$\mu$m images, and the bottom row the SPIRE~500~$\mu$m and
  a NASA/JPL-Caltech composite image. The colorbar under the
  {\em Herschel} images shows the flux density in Jy/pixel. The grayscale
  image is the NASA/JPL-Caltech composite image (pia15817.html)
  including infrared data from Spitzer wavelengths 3.6 to 4.5~$\mu$m
  and 8 to 24~$\mu$m, WISE data at 3.4 to 4.5 and 12 to 22~$\mu$m, and
  ultraviolet data from GALEX at 0.15 to 2.3~$\mu$m. The region
  observed by {\em Herschel} and considered in this paper is indicated by
  the white contour in this last image.}
\label{imN7293}
\end{center}
\end{figure*}

To generate broadband images, we used the {\em Herschel} Interactive
Processing Environment (HIPE, version 10.0; \cite{Ott10}) and the
Scanamorphos data reduction tool (Scanamorphos, version 20;
\citet{Roussel13}). We used the same pixel size for the PACS 70 and
160 $\mu$m images being 2\farcs85, and the standard Scanamorphos pixel
sizes of 4\farcs5, 6\farcs25, and 9\arcsec\, for the SPIRE images. The
FWHM of the PACS beam at 70 and 160~$\mu$m is 5\farcs7 and 11\farcs4,
respectively. The FWHM of the SPIRE beam sizes are 18\farcs1,
25\farcs2, and 36\farcs6 at 250, 350, 500~$\mu$m, respectively. The
PACS images are in Jy/pixel, while SPIRE images are in Jy/sr and were
converted to Jy/pixel using the above pixel and beam sizes within
HIPE. The background sources outside the Helix and known sources
across the Helix were subtracted using SExtractor and the images were
sky subtracted within HIPE. It is possible that some contaminating
hitherto unknown sources (mainly background galaxies) remain across
the nebula. The resulting images are presented in Fig.~\ref{imN7293}.

\begin{table*}
\caption{The photometric imaging observation log of the Helix nebula with {\em Herschel}. }
\begin{tabular}{rrrlrrr}
\hline
OD & RA (J2000.0) & DEC(J2000.0) & AOT & Duration & Start Date \& Time & Obsid \\
\hline
350  & 22h29m38.55s & -20d50m13.60s & SpirePacsParallel   & 4326s & 2010-Apr-28T22:30:39 & 1342195681 \\
350  & 22h29m38.55s & -20d50m13.60s & SpirePacsParallel   & 4116s & 2010-Apr-28T23:43:58 & 1342195682 \\
1304 & 22h29m35.24s & -20d50m40.51s & SpirePhotoLargeScan & 2047s & 2012-Dec-08T10:54:32 & 1342256744 \\
\hline
\end{tabular}
\label{tabObs}
\end{table*}

\section{Dust morphology}
\label{sectMorph}

Only a bit of the outer halo and arc in the north is detected. Hence
we will restrict the discussion in this paper to the main nebula. The
region considered is indicated on the grayscale image in
Fig.~\ref{imN7293}.

\citet{Zeigler13} noted that the structure of the Helix projects as if
it were a thick walled {\it barrel} composed of red- and blue-shifted
halves in a bipolar geometry. The barrel axis of the Helix is tilted
about 10$\degr$ east and 6$\degr$ south relative to the line of sight
\citep{Zeigler13}. The barrel rims are widely separated in space, but
the projected barrel shape creates a false impression of the inner
and outer circular structures. The apparent so-called ``inner ring''
is actually spatially distant velocity components projected together
to create the illusion of a circle. Similarly, the apparent so-called
{\it outer ring} is an artifact of the bulging central wall of the
barrel, pushed outward to the east and west.

In the SPIRE maps we look at the cold dust without emission from
atomic ions. The best compromise of resolution and signal is provided
by the SPIRE 250~$\mu$m MD map (Fig. \ref{imN7293}). The barrel wall
is very fragmented and consists of emission clumps on size scales down
to the resolution of the map. In the N we look more at the outer edge
of the barrel, while in the S we look more down the barrel towards the
central star. The dust emission is fainter when we look inside the
barrel, and brighter when we look at the outside of the barrel wall.
There appears to be less dust in the NW, which is in the direction of
the receding plume. These plumes in the NW and SE appear as a radial
flow-like structure in the WISE and GALEX images (see grayscale image
in Fig.~\ref{imN7293}). They may be caused by a bipolar outflow with
somewhat higher velocity puncturing wholes in the barrel walls
\citep{Zeigler13}. The plume in the SE is broader but less obvious,
because we are looking inside, down the barrel towards the central
star. The inner and outer side of the barrel appear well separated
except in the SSW and in the N. In these regions the dust emission is
bright, as well as in two extensions in the NNW and SSE. The latter are
the edges of the barrel lobes.

Adopting the thick walled barrel bipolar morphology, as proposed by
\citet{Zeigler13} and \citet{Meaburn08}, we notice that the dust emission
comes mainly from the barrel wall. The dust emission is
strongest where we look from the outside at the barrel, 
and is weaker when we look inside the barrel towards the
central star.

\section{Spectral energy distribution}
\label{sectSED}

\begin{figure}
\begin{center}
\includegraphics[width=\columnwidth,keepaspectratio]{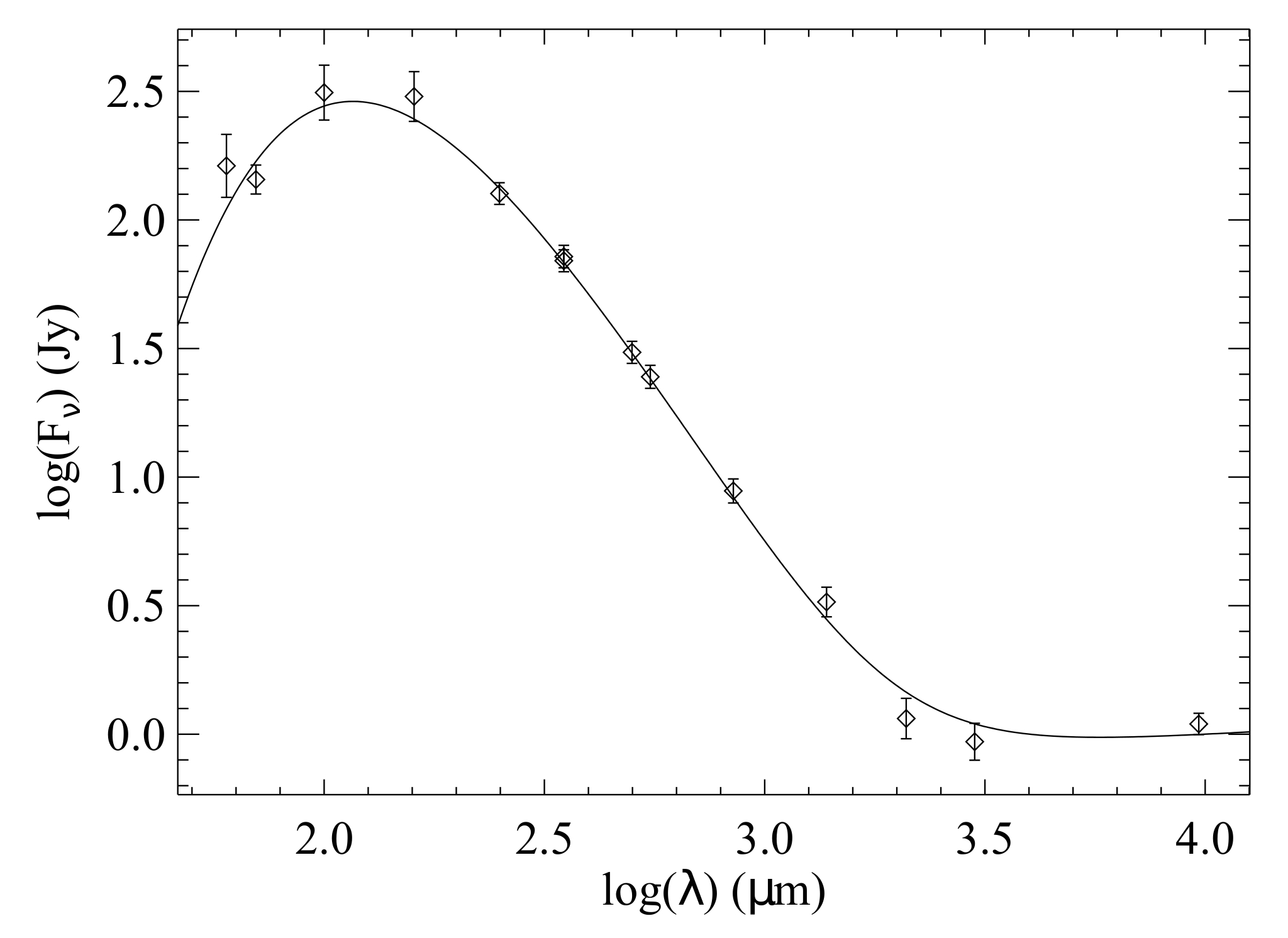} 
\caption{The SED of NGC~7293. IRAS, {\em Herschel} PACS and SPIRE, Planck, and CBI interferometer data points are fit by a modified blackbody as in Eq.~\ref{equ2} with $\beta$=0.99 and T$_{dust}$$=$30.8~K}
\label{figSED}
\end{center}
\end{figure}

\begin{table}
\caption{ 
The far infrared continuum fluxes measured by various satellite
missions and the CBI interferometer. The second column gives the
wavelength, the third the quoted flux, the fourth column the
conversion factors from point source calibration to extended source
calibration for the SPIRE data, the fifth column the color correction
factor, the sixth column the actual flux with all corrections applied,
the seventh column the estimated error on this flux, and the eighth column the flux value from the blackbody fit.
}
\begin{tabular}{lrrrrrr}
\hline
Instr & $\lambda$  & K$_{4E}$/ & K$_{color}$ & flux     & error   & fit \\
           & $\mu$m     &    K$_{4P}$ &           & Jy       & Jy      & Jy  \\
\hline
 IRAS &   60  &  1.0000 & 0.9457 & 162.3 &   52.9 &   92.6 \\ 
 IRAS &  100   &  1.0000 & 0.9650 & 312.8 &   87.0 &  271.9 \\
 PACS &   70   &  1.0000 & 0.9961 & 143.6 &   19.9 &  151.8 \\
 PACS & 160   &  1.0000 & 0.9996 & 302.0 &   75.3 &  251.2 \\
SPIRE & 250   &  0.9828 & 0.9978 & 126.7 &   12.9 &  135.6 \\
SPIRE & 350   &  0.9843 & 0.9991 & 69.4  &    7.2 &   69.6 \\
SPIRE & 500   &  0.9710 & 0.9788 & 30.6  &    3.2 &   30.8 \\
Planck & 350  &  1.0000 & 1.0190 & 72.0  &    7.6 &   69.6 \\
Planck & 550  &  1.0000 & 1.0881 & 24.5  &    2.7 &   24.5 \\
Planck & 849  &  1.0000 & 1.0983 &  8.83  &    1.00 &    8.42 \\
Planck & 1382 &  1.0000 & 1.1016 &  3.27  &    0.46 &    2.81 \\
Planck & 2096 &  1.0000 & 1.0077 &  1.15  &    0.23 &    1.46 \\
Planck & 2998 &  1.0000 & 1.0695 &  0.93  &    0.17 &    1.10 \\
CBI & 9671 &  1.0000 & 1.0000 &  1.10  &    0.11 &    1.00 \\
\hline
\end{tabular}
\label{tabSED}
\end{table}

In Fig.~\ref{figSED} we present the spectral energy distribution (SED)
of the Helix nebula. We retrieved the Planck images from the Planck
database and converted the units to Jy/pixel (see also
\citet{Planck14}). We determined the integrated flux down to 0.008
Jy/pixel in the Planck 350~$\mu$m image, which corresponds to about
3~$\sigma$. We used this contour to measure the flux values at all
other Planck wavelengths at 350, 550, 849, 1382, 2096, 2998~$\mu$m,
but also used the same region within this same contour to measure the
flux values in the SPIRE 250, 350, 500 $\mu$m parallel mode images,
and the PACS 70 and 160~$\mu$m parallel mode images convolved to the
SPIRE 250~$\mu$m beam and rebinned to its 4\farcs5 pixel size with flux
conservation. We also retrieved the IRAS 60 and 100~$\mu$m images, but
used a different contour as the IRAS beams were too large and too much
flux would have been missed. We added a point at 31 GHz (9671~$\mu$m)
from \citet{Cassasus04} obtained with the Cosmic Background Imager (CBI)
interferometer in Chile \citep{Padin02} to constrain the free-free
emission. Photometric color corrections were applied to all flux
densities. This correction is needed to convert monochromatic flux
densities that refer to a constant energy spectrum to the true object
SED flux densities at the photometric reference wavelengths of each
instrument. The data are summarized in Table~\ref{tabSED}. Since the
source is extended, the appropriate correction for extended source
calibration were applied to the SPIRE fluxes
\begin{equation}
F_\nu{\rm [actual]} = \frac{K_{\rm 4E}}{K_{\rm 4P}}\frac{F_\nu{\rm [quoted]}}{K_{\rm color}},  
\end{equation}
The correction factors are given in Table~\ref{tabSED}. We note that the
definition of $K_{\rm color}$ is the inverse of the definition used in the
SPIRE Observer's Manual, but agrees
with the definition used by the other instrument teams.

The Planck fluxes are a bit lower, but up to 2096~$\mu$m (143~GHz)
within the error bars of the flux values mentioned in
\citet{Planck14}. This is because the contours have been optimized
for the SPIRE images where the beam and pixel sizes are smaller and the region
considered more adjusted to the nebula. The
percentage differences in flux values at longer wavelengths are larger,
because the flux values are low and the background more significant
compared to the shorter wavelengths (higher frequencies).

The thermal radiation emitted from a population of dust particles
depends on their temperature distribution and opacity. Assuming that
the opacity index varies with frequency as a power law, 
$\kappa_{\nu}=\kappa_{0}(\nu/\nu_{0})^{\beta}$, and that the dust
grains dominating the flux in the IR and submm part of the SED all
have the same temperature $T_{\rm dust}$, and the dust emission is
optically thin, then the flux density spectral distribution can be
approximated with a modified blackbody function of that temperature,
plus a contribution of free-free emission given by
\begin{equation}                              
F_\nu = C \nu^\beta B_\nu(T_{\rm dust}) + S_{1GHz}*(\nu/1~GHz)^{-0.1},  
\label{equ2}
\end{equation}
where $\nu$ is the frequency and B$_{\nu}$ the Planck function of the
dust temperature T$_{\rm dust}$, and S$_{1GHz}$~$=$~1.4~$\pm$0.4~Jy.
The power law exponent of the opacity law, $\beta$, basically depends
on material properties and the size distribution of the dust
grains. An exponent ${\beta}~\approx~0$ would point to grains
radiating almost as blackbodies \citep{Hildebrand83}.

We made $\chi^{2}$-minimization fits to the available photometry data
weighted with the accuracy of the flux density measurements and a
contribution for the free-free emission. We obtained a dust
emissitivity index $\beta$$=$0.99~$\pm$~0.09 and a dust temperature
T$_{\rm dust}$$=$30.8~K~$\pm$1.4~K. Both measured and fitted values
are presented in Table \ref{tabSED} and the best fit model in
Fig.~\ref{figSED}. A dust emissivity of slightly less than 1.0 was
also found in both C-rich and O-rich (post-)AGB stars. This indicates
that the dust has a layered amorphous structure \citep{Knapp93}. In
laboratory measurements a $\beta$ value of 1 is more typical for amorphous
carbon grains than for silicate grains \citep{Mennella95, Mennella98}.
The Planck collaboration found a temperature of 34~K (no error bar
given) and an amorphous carbon grain composition based on modeling
the SED with DUSTY \citep{Planck14}. DUSTY assumes spherical symmetry
for the dusty envelope and the authors used a r$^{-2}$ density
distribution as first approximation to the nebular parameters.

The composition of the progenitor star in the Helix is still a subject
of debate. Measurements by \citet{Henry99} of atomic emission lines in
the nebula at one position in the northeastern rim indicate that
C/O is 0.87~$\pm$~0.12 suggesting that hot-bottom burning on the late
AGB converted much of the $^{12}$C into $^{14}$N. However, the
presence of CN, HCN, HNC, c-C$_3$H$_2$, and C$_2$H measured at one
position in the eastern part of the Helix indicates a C-rich
environment \citep{Tenenbaum09}. In view of the carbon richness of
the Helix molecular envelope, the absence of the usual PAH emission
features in the ISOCAM CVF \citep{Cox98} and IRS spectra
\citep{Hora06} spectra is surprising. The absence of PAH emission
bands in the mid-infrared spectra implies that no small dust particles
are present in the envelope of this evolved PN \citep{Cox98} as they may have
been destroyed during PN evolution, or it may
be a sensitivity effect \citep{Hora06}. The fact that dust grains are
larger than typical grains in the ISM has been seen in other PNe 
such as NGC~650 \citep{vanHoof13} and NGC~6445 \citep{vanHoof00}
for example. Several explanations were proposed in \citet{vanHoof13},
but the most plausible explanation could be that the large grains
formed already during the AGB stage.

\section{Temperature map}
\label{sectTemp}

\begin{figure}
\begin{center}
\includegraphics[width=0.98\columnwidth,keepaspectratio]{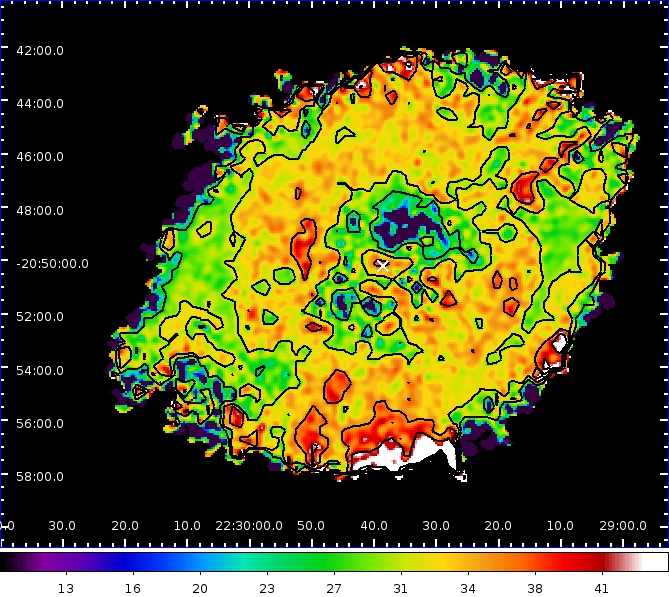} 
\caption{Temperature map of NGC~7293 based on convolved PACS~70~$\mu$m 
and SPIRE~250~$\mu$m~MD images. The temperature scale is shown below and the
contours are at 22, 31, and 35~K. }
\label{imtemp}
\end{center}
\end{figure}

The PACS 70~$\mu$m sky subtracted image was convolved to the SPIRE
250~$\mu$m beam using the appropriate convolution kernels of
\citet{Aniano11} with flux conservation and rebinned to the pixel size
of 4\farcs5. We computed the ratio of this convolved PACS 70~$\mu$m
image to the SPIRE 250~$\mu$m MD image.  We computed a table of
70~$\mu$m/250~$\mu$m flux ratios of a modified black body at different
temperatures with $\beta$$=$1.0 folded with the PACS and SPIRE filter
transmission curves available in HIPE (v.8.0.1387) using the
procedure outlined in the SPIRE Observer's Manual. Using this table we
interpolated the flux ratios in the ratio image mentioned above as a
function of temperature at each pixel to obtain the temperature map
(Fig.~\ref{imtemp}). The central star position is marked with a white
cross. We clipped the temperature image to the 1$\sigma$ contour of
the sky subtracted SPIRE 250~$\mu$m image.

The dust temperature is generally between about 22~K and 42~K. The
dust is warmest in low density regions irradiated directly by the
central star, and is cooler in dense regions and regions farther away
from the central star and/or shielded from direct starlight. We
notice a warmer ring-like structure. This ring-like structure
encompasses the inner side of the barrel as defined by
\citet{Zeigler13}. The temperature in the barrel wall is not at all
uniform but reflects its clumpiness. The extensions in the SE and NW
are cool, high opacity regions at the outside of the barrel farther
away from the central star and shielded from direct starlight. Towards
the plumes in the NW there are clumps of higher temperature. The
higher temperatures in the S are mainly due to the excess emission in
the PACS 70~$\mu$m map. As we are looking down the inside of the
barrel towards the central star, it may well be that the temperature
is higher. At other wavelengths there are also indications of a
higher temperature in this region. Because of the low signal-to-noise
ratio in the 70~$\mu$m map and the lower emission in the SPIRE
250~$\mu$m map at the southern edge, we are not very sure of exactly
how warm this region would be. It is reasonable to assume that it is
not much warmer than other regions in the center of the nebula, where
the temperature is about 40~K.

\section{Dust mass column density }
\label{sectMass}

\begin{figure}
\begin{center}
\includegraphics[width=\columnwidth,keepaspectratio]{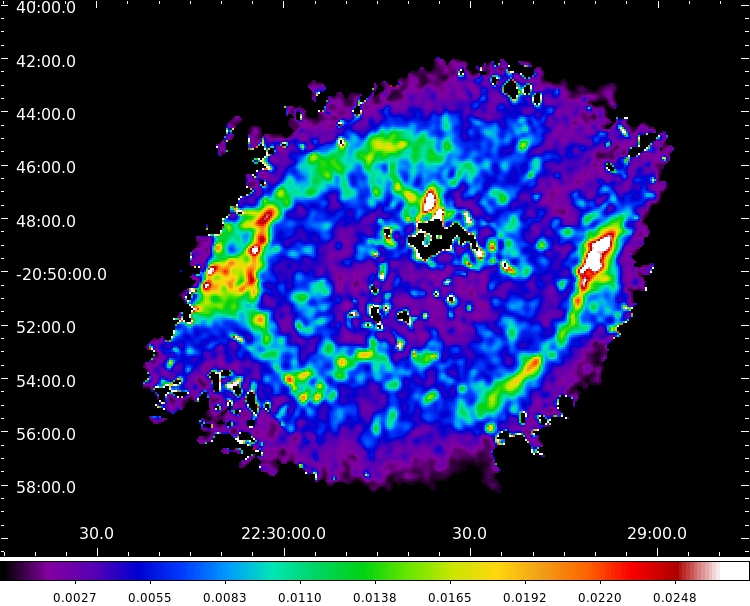} 
\caption{Dust mass column density map of NGC~7293 based on SPIRE~250~$\mu$m~MD image 
and the temperature map. The mass density scale is shown below in M$_\odot$/pc$^2$. }
\label{imMass}
\end{center}
\end{figure}

We calculated the dust mass column density for each pixel based on the
temperature and the SPIRE~250~$\mu$m~MD image using the formula of
\citet{Gledhill02} and \citet{Hildebrand83} (Fig.~\ref{imMass}). The
dust mass density image is not uniform, but reflects the clumpiness of
the Helix nebula. The densest regions are regions where we look at
the edge of the barrel. Integrating values in the map and assuming a
distance of 216~pc the total dust mass of the Helix is 3.5~10$^{-3}$
M$_\odot$.

As in \citet{Gledhill02} we have assumed a bulk density of
$\rho_{gr}~=~3~g~cm^{-3}$. According to the \citet{Planck14} paper
this value would be more appropriate for silicate grains. Because it
appears that amorphous carbon would be most appropriate for the dust
composition, they used a value of $\rho_{gr}~= 2~g~cm^{-3}$. They also
used a distance of 213~pc instead of 216~pc. When we use these values
we find a total dust mass of 3.4~10$^{-3}$~M$_\odot$, which is in
agreement with their value of 3.6~10$^{-3}$~M$_\odot$ from their DUSTY
modeling.

When we compare the dust temperature map Fig.~\ref{imtemp} with the
mass column density Fig.~\ref{imMass} map we notice that the dust
temperature is about 31~K where the mass density is highest, and above
38~K where the density is lower.

\section{Properties of the central star disk }
\label{sectDisk}

\begin{figure}
\begin{center}
\includegraphics[width=0.90\columnwidth,keepaspectratio]{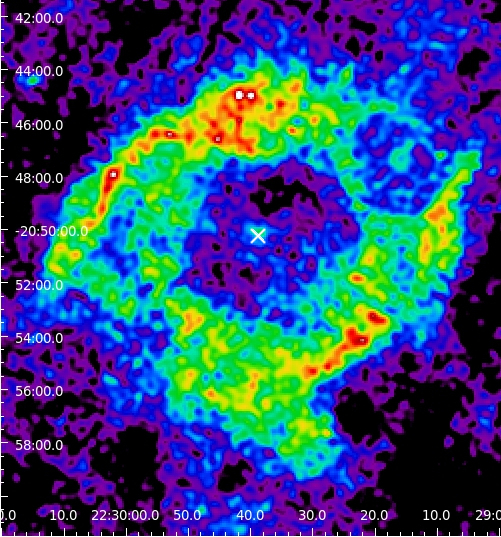}
\caption{PACS $70$ image convolved to the SPIRE~$250$~$\mu$m
beam with the central star disk indicated with a white cross.}
\label{imPACSdisk}
\end{center}
\end{figure}

\citet{Su07} detected a pointlike source coincident with the central
star, and a plateau of diffuse emission in the 24~$\mu$m image, but
only a pointlike source of flux 224~$\pm$~33~mJy at 70~$\mu$m, and no
point-like or extended emission at 160~$\mu$m with Spitzer.  We
convolved the PACS images to the SPIRE 250~$\mu$m beam using a pixel
size of 4\farcs5 (see Fig.~\ref{imPACSdisk}) and also noticed the
presence of the central source at 70~$\mu$m, with a flux of
239~$\pm$~24~mJy (in good agreement with Su et al.), but not at
longer wavelengths, and no diffuse emission at any
wavelength. Combining the 24 and 70~$\mu$m flux and our upper limits
at other {\em Herschel} wavelengths would imply a temperature of at
least 65~K for the debris disk, and a mass of around
2~10$^{-7}$~M$_\odot$, with a large degree of uncertainty due to the four
upper limits used. 

\section{Comparison with images at other wavelengths}
\label{sectComp}

\subsection{Comparison with GALEX images}

\begin{figure}
\begin{center}
\includegraphics[width=0.98\columnwidth,keepaspectratio]{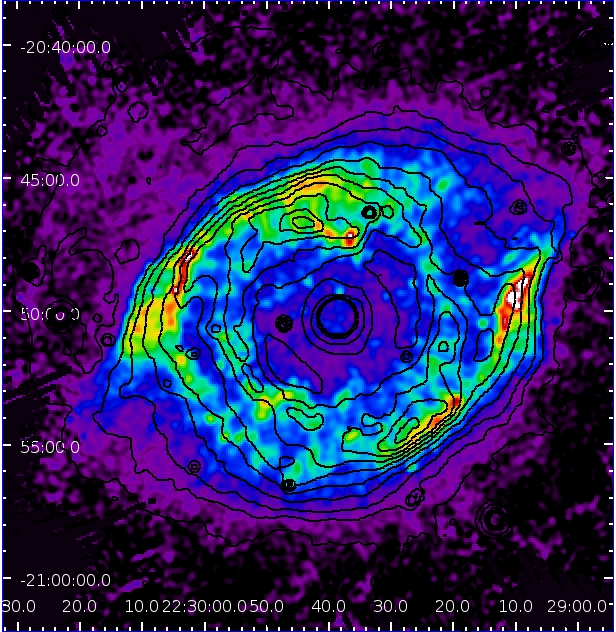} 
\caption{SPIRE 250~$\mu$m MD image with the contours of GALEX NUV image convolved 
to SPIRE 250~$\mu$m beam overlaid. }
\label{N7293SP250GAL}
\end{center}
\end{figure}

\begin{figure}
\begin{center}
\includegraphics[width=0.98\columnwidth,keepaspectratio]{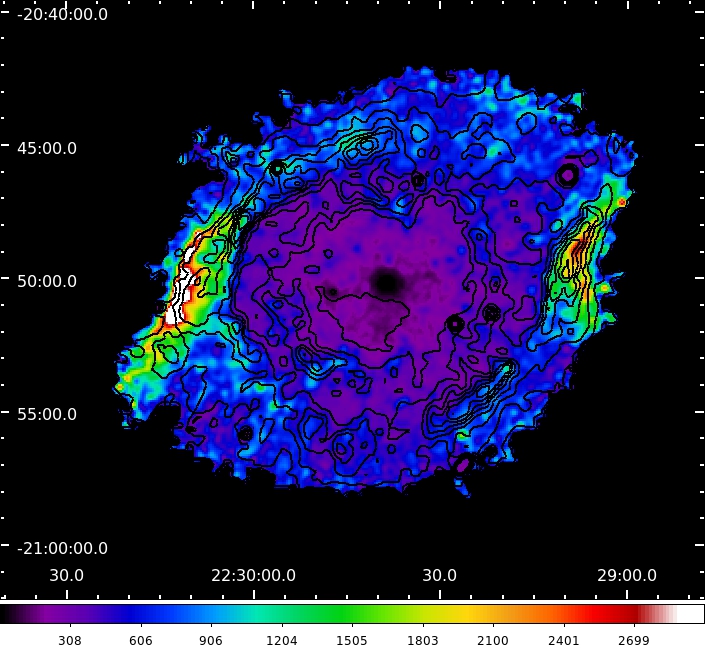} 
\caption{The ratio of the SPIRE 250~$\mu$m image to the NUV GALEX image convolved to the SPIRE 250~$\mu$m beam with the H$_2$~1$-$0~S(1) contours of \citet{Speck02} image convolved to the SPIRE 250~$\mu$m beam overplotted.}
\label{im_ratioSP250NUV_H2cont}
\end{center}
\end{figure}

In Fig.~\ref{N7293SP250GAL} we overlay the contours of the NUV (175~$-$~
280~nm) GALEX image, retrieved from the GALEX archive and convolved to
the SPIRE 250~$\mu$m beam, on the SPIRE 250~$\mu$m MD image. There is
UV emission to the same spatial extent as the dust emission, but we
see that the brightest dust emission is located just outside the
bright emission in the NUV image. The UV emission is strongest at the
edge of the inner cavity while the far-infrared emission comes mainly
from the outer edge of the barrel.

In Fig.~\ref{im_ratioSP250NUV_H2cont} we show the ratio of this
convolved NUV GALEX image to the SPIRE 250~$\mu$m MD image.  As mentioned
before this image emphasizes that the dust emission is clumpy
and the NUV emission is passing in between the clumps. The dust
emission is strongest relative to the NUV emission in the extension in
the SE. It is remarkable as the dust emission in the SPIRE 250~$\mu$m
map nor the dust density seems particularly strong in this
region. Here the dust appears to be shielded most from the UV
emission. We are looking face on at the outer part of the barrel
wall. This extension is moving towards us in the HCO$^+$ map of
\citet{Zeigler13}. The extension in the NW has more complex kinematics
and in part is moving away from us. It is the projection of the rim of
the barrel lobe more towards the back of the Helix. It must be
illuminated by the radiation passing through the holes. For the region
in the south where there is relatively little dust emission compared
to the NUV emission, we are looking down the barrel with blueshifted
HCO$^+$ velocities in front and redshifted velocities towards the back
of the barrel. This is the region where there seems to be an excess in
70~$\mu$m emission (Fig. \ref{imN7293}), a somewhat higher temperature
(Fig. \ref{imtemp}) and on average lower dust column density
(Fig. \ref{imMass}).

In summary, the dust emission is strongest compared to the NUV
emission where we look at the outside of the barrel of the Helix and the
dust emission is most shielded from the UV radiation. The dust is 
clumpy and UV radiation must leak through.

\subsection{Comparison with the H$_2$ image}
\label{compH2}

\begin{figure}
\begin{center}
\includegraphics[width=0.98\columnwidth,keepaspectratio]{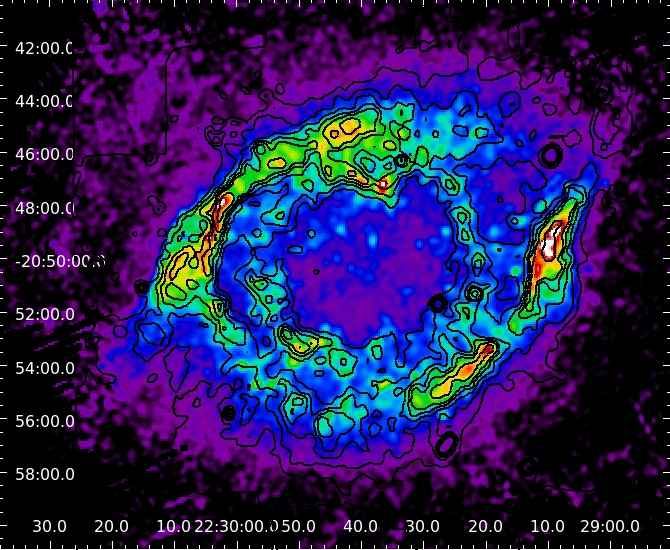} 
\caption{The contours of the H$_2$~1$-$0 S(1) image convolved to the SPIRE~250~$\mu$m beam 
are overlaid on the SPIRE~250~$\mu$m image}
\label{imH2}
\end{center}
\end{figure}

\begin{figure}
\begin{center}
\includegraphics[width=0.98\columnwidth,keepaspectratio]{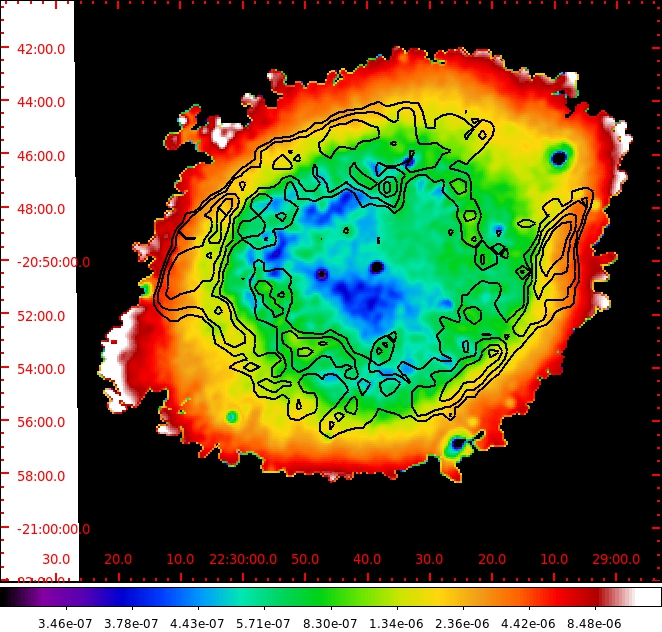} 
\caption{The ratio of the H$_2$ to the H${\beta}$ image both convolved to the 
SPIRE 250~$\mu$m beam (displayed in log scale) with
 the dust SPIRE 250~$\mu$m contours overlayed.}
\label{im_ratioH2_Hbeta}
\end{center}
\end{figure}

We convolved the H$_2$ image of \citet{Speck02} with the SPIRE
250~$\mu$m PSF \citep{Aniano11}. The contours of this convolved H$_2$
image are overlaid on the SPIRE 250~$\mu$m image Fig.~\ref{imH2}. We
see that the H$_2$ emission comes from the same region as the dust
emission. The peaks of the H$_2$ emission and the dust emission agree,
except in the south. The H$_2$ emission is relatively strong compared
to the dust emission in the south and the peaks do not seem to
coincide. This is the region where we look down the barrel towards the
central star.

From high resolution H$_2$ images it was found that in the Helix the
H$_2$ emission is always associated with dense clumps, {\it knots}, that
are embedded in the ionized gas \citep{Matsuura09}. The gas kinetic
temperature T$_k$ was determined to be about 20 to 40~K by
\citet{Zack13} and \citet{Etxaluze14}, which is similar to our range
in dust temperatures determined in Sect.~\ref{sectTemp}. The gas
density of the H$_2$ cometary globules is on the order of
n(H$_2$)~$\sim$~(1-5)~10$^5$ cm$^{-3}$ \citep{Zack13, Etxaluze14}.
\citet{Goldsmith01} found that for gas densities higher than
10$^{4.5}$~cm$^{-3}$ the dust and gas temperatures will be closely
coupled, also for temperatures as found in the Helix. It is another
indication that, as is the case in the Ring Nebula \citep{vanHoof10}
and NGC~650 \citep{vanHoof13}, also in NGC~7293 H$_2$ and dust
co-exist in these knots. This close correlation could be an indication
of the formation of H$_2$ on dust grains.

Exactly when the knots formed during the evolution is still debated.
These knots may be density enhancements formed during the AGB phase in
the shell and survived the ionization of the nebula
\citep{Matsuura09,Aleman11}, or they may have formed because of
instabilities and fragmentation at the ionization front when the
nebula got ionized \citep{Garcia-Segura06, Huggins06}. As knots are
observed in older PNe having central stars well evolved on the cooling
track, it has been suggested that they formed because of hydrodynamical
instabilities when the nebular gas is going through recombination
\citep{ODell07,vanHoof10}. The formation of H$_2$ most effectively
takes place on the surface of dust grains, also at grain temperatures
as found here for the Helix nebula \citep{Cazaux02,Cazaux04}. The
high density in the knots makes it possible to shield the H$_2$ from
the stellar radiation field. The high densities and low dust
temperatures observed allow H$_2$ molecules to be reformed on dust grains
on a reasonable timescale \citep{vanHoof10}. Molecular hydrogen formation in the
knots is expected to be substantial after the central star entered the
cooling track and underwent a strong drop in luminosity, and may still
be ongoing at this moment, depending on the density of the knots and
the properties of the grains in the knots and assuming the knots
formed quickly. Especially in the latter case we expect a tight
correlation between the dust emission and the H$_2$ emission
\citep{vanHoof10} as is observed in the Helix nebula.

The possible effect of the UV emission on the H$_2$ is investigated in
Fig.~\ref{im_ratioSP250NUV_H2cont}. H$_2$ contours are overlaid on the
ratio of the SPIRE~250~$\mu$m image to the NUV Galex image. We see
that the strong dust and H$_2$ emission originates just outside the
peaks of NUV emission at the edges of the nebula and in the
extensions. It is in these regions where the H$_2$ and dust emission
coincides best.  When we overplot the H$_2$ contours on the ratio
image of the SPIRE 250~$\mu$m MD to NUV images, we notice that the
H$_2$ emission is situated at the inner edge where this ratio is very
high. This is in agreement with what was found by \citet{Aleman11}: if
the cometary knot is beyond the Helix ionization front, then there is
no ionized region, the knot is completely neutral and there would be
not enough radiation or temperature to excite significantly the upper
vibrational levels of the molecule. The 1–0 S(1) intensity would be
very low. In the south where there is relatively more NUV flux to dust
emission the H$_2$ emission is relatively strong.

To compare the molecular with the ionized component we obtained an
H$\beta$ image from \citet{Corradi03} and \citet{ODell98}. We
convolved the image to the SPIRE 250~$\mu$m beam \citep{Aniano11} with
flux conservation and rebinned it to a pixel size of 4\farcs5. We
took the ratio of the convolved H$_2$ image to this H$\beta$ image.
The resulting image is shown in Fig.~\ref{im_ratioH2_Hbeta}. The
density structure has virtually disappeared and we see the rapid
decrease of the ionizing radiation field outwards beyond the inner
ring. We see that the ionized region is mainly contained within the
molecular and dust region. The dust and molecular emission is not
particularly strong inside the barrel, because the influence of the
stellar radiation field is significant.

\section{Conclusion}
\label{sectConcl}

We presented {\em Herschel} PACS and SPIRE images of the Helix nebula. The
dust emission is clumpy and predominantly present in the barrel wall. We presented
consistent photometry of the Helix without its halo and determined its
spectral energy distribution. The emissivity index of $\beta = 0.99
\pm 0.09$, in combination with the carbon rich molecular chemistry of
the nebula, indicates that the dust consists mainly of amorphous
carbon. We determined an average dust temperature of 30.8~K $\pm$ 1.4.
We detected the central star disk at 70~$\mu$m, confirming the result
of \citet{Su07}. At other {\em Herschel} wavelengths
we only have upper limits, hence the dust temperature of the disk of
65~K is only an estimate. The dust mass column density map reflects
the clumpiness of the dust. The dust mass of the main nebula is 
$\sim$~3.5~10$^{-3}$~M$_\odot$ at a distance of 216~pc. The temperature map
shows a variation in dust temperature between 22~$-$~42~K. This is
similar to the kinetic temperature of the molecular gas and indicates that the
dust and molecular gas co-exist in the dense clumps. We compared the
SPIRE 250~$\mu$m image with images at other wavelengths to determine
the extent and morphology of the dust, molecular, and ionized
component in the nebula. The morphology of the different components
can be understood adopting the thick walled barrel-like model tilted
about 10\degr\ east relative to the line of sight from
\citet{Zeigler13}. The radiation field decreases rapidly outwards in
the barrel wall. The very good coincidence between the H$_2$ and the
dust emission in the barrel wall suggests the formation of H$_2$ on
dust grains.

\begin{acknowledgements}

  {\em Herschel} is an ESA space observatory with science
  instruments provided by European-led Principal Investigator
  consortia and with important participation from NASA. 
  PACS has been developed by a consortium of institutes led by MPE
  (Germany) and including UVIE (Austria); KU Leuven, CSL, IMEC
  (Belgium); CEA, LAM (France); MPIA (Germany); INAF-IFSI/OAA/OAP/OAT,
  LENS, SISSA (Italy); IAC (Spain). This development has been
  supported by the funding agencies BMVIT (Austria), ESA-PRODEX
  (Belgium), CEA/CNES (France), DLR (Germany), ASI/INAF (Italy), and
  CICYT/MCYT (Spain)."

  SPIRE has been developed by a consortium of institutes led by
  Cardiff University (UK) and including Univ. Lethbridge (Canada);
  NAOC (China); CEA, LAM (France); IFSI, Univ. Padua (Italy); IAC
  (Spain); Stockholm Observatory (Sweden); Imperial College London,
  RAL, UCL-MSSL, UKATC, Univ. Sussex (UK); and Caltech, JPL, NHSC,
  Univ. Colorado (USA). This development has been supported by
  national funding agencies: CSA (Canada); NAOC (China); CEA, CNES,
  CNRS (France); ASI (Italy); MCINN (Spain); SNSB (Sweden); STFC (UK);
  and NASA (USA)

  HCSS / HSpot / HIPE is a joint development (are joint developments)
  by the {\em Herschel} Science Ground Segment Consortium, consisting of
  ESA, the NASA {\em Herschel} Science Center, and the HIFI, PACS and SPIRE
  consortia.

  This research made use of tools provided by Astrometry.net. 

  Some of the data presented in this paper were obtained from the
  Multimission Archive at the Space Telescope Science Institute
  (MAST). STScI is operated by the Association of Universities for
  Research in Astronomy, Inc., under NASA contract NAS5-26555. Support
  for MAST for non-HST data is provided by the NASA Office of Space
  Science via grant NNX09AF08G and by other grants and contracts.

  P.v.H. and the PACS ICC in Leuven wish to acknowledge support from
  the Belgian Science Policy office through the ESA PRODEX programme.

  ME, JRG and JC thank the Spanish MINECO for funding support from
  grants CSD2009-00038, AYA2009-07304 and AYA2012-32032.

\end{acknowledgements}

\bibliographystyle{aa} 
\bibliography{n7293}

\end{document}